\documentclass[
twocolumn,
prx,showpacs,preprintnumbers,amsmath,amssymb]{revtex4}

\usepackage{times}
\usepackage{graphicx}
\usepackage{amssymb}
\usepackage{amsthm}
\usepackage{amsmath}
\usepackage{dsfont}
\usepackage{bm}
\usepackage{mathrsfs}
\usepackage{bbold}

\def\Xint#1{\mathchoice
{\XXint\displaystyle\textstyle{#1}}%
{\XXint\textstyle\scriptstyle{#1}}%
{\XXint\scriptstyle\scriptscriptstyle{#1}}%
{\XXint\scriptscriptstyle\scriptscriptstyle{#1}}%
\!\int}
\def\XXint#1#2#3{{\setbox0=\hbox{$#1{#2#3}{\int}$ }
\vcenter{\hbox{$#2#3$ }}\kern-.6\wd0}}

\def\dashint{\Xint-}

\begin{document}
\title{Classical analogue of the Unruh effect}
\author{Ulf Leonhardt$^1$, Itay Griniasty$^1$, Sander Wildeman$^2$, Emmanuel Fort$^2$, and Mathias Fink$^2$}
\affiliation{
$^1$Weizmann Institute of Science,
Rehovot 761001,
Israel\\
$^2$Institut Langevin, ESPCI, CNRS, PSL Research University, 1 rue Jussieu, 75005 Paris, France
}
\date{\today}
\begin{abstract}
In the Unruh effect an observer with constant acceleration perceives the quantum vacuum as thermal radiation. The Unruh effect has been believed to be a pure quantum phenomenon, but here we show theoretically how the effect arises from the classical correlation of noise. We demonstrate this idea with a simple experiment on water waves where we see the first indications of a Planck spectrum in the correlation energy. 
\end{abstract}

\maketitle

\section{Introduction}

Imagine an observer moving through the quantum vacuum of empty space. In free space, the quantum vacuum is Lorentz invariant, so a uniformly moving observer would not see any effect due to motion, but an accelerated observer would. This is known as the Unruh effect \cite{Unruh} (or Fulling--Davies--Unruh effect in full \cite{Unruh,Fulling,Davies}). An observer with constant acceleration $a$ is predicted \cite{Unruh} to perceive empty space as thermal radiation with Unruh temperature
\begin{equation}
K_\mathrm{B}T = \frac{\hbar a}{2\pi c} 
\label{unruh}
\end{equation}
where $c$ is the speed of light in vacuum, $\hbar$ Planck's constant divided by $2\pi$ and $K_\mathrm{B}$ Boltzmann's constant.

The Unruh effect and the closely related Bekenstein-Hawking radiation of black holes \cite{Bekenstein,Hawking} has been one of the most important results of theoretical physics of the second half of the 20th century, hinting on a hidden connection between three vastly different areas of physics indicated by the constants appearing in Eq.~(\ref{unruh}): general relativity (acceleration $a$ versus $c$), quantum mechanics ($\hbar$) and thermodynamics ($K_\mathrm{B}$). It has been the benchmark for theories attempting to unify these areas ever since.

Yet there has been no experimental evidence for the Unruh effect. The reason becomes evident if one puts numbers into Unruh's formula: with $\hbar\approx 10^{-34}\mathrm{Js}$ and $c\approx 3\times10^8\,\mathrm{m}/\mathrm{s}$ one needs an acceleration of about $10^{23}\,\mathrm{m}/\mathrm{s}^2$ to reach room temperature. Three avenues \cite{Crispino} have been suggested for getting closer to an observation of Unruh radiation: {\bf i)} strong-field acceleration such as in laser plasmas, wakefields or strongly accelerated electrons, {\bf ii)} cavity QED and {\bf iii)} particle accelerators; none have been successful so far. 

\begin{figure}[t]
\begin{center}
\includegraphics[width=11.0pc]{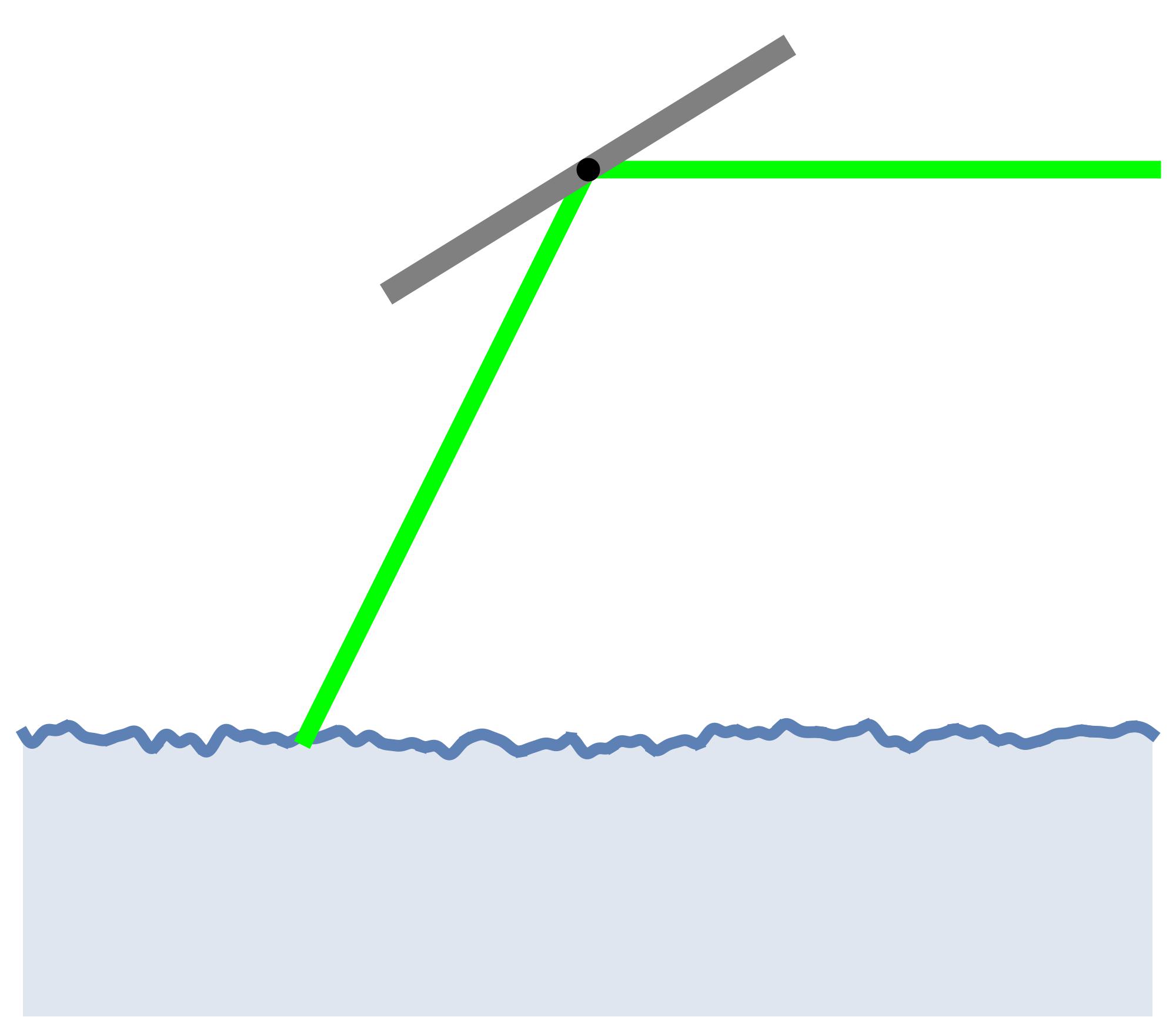}
\includegraphics[width=15.5pc]{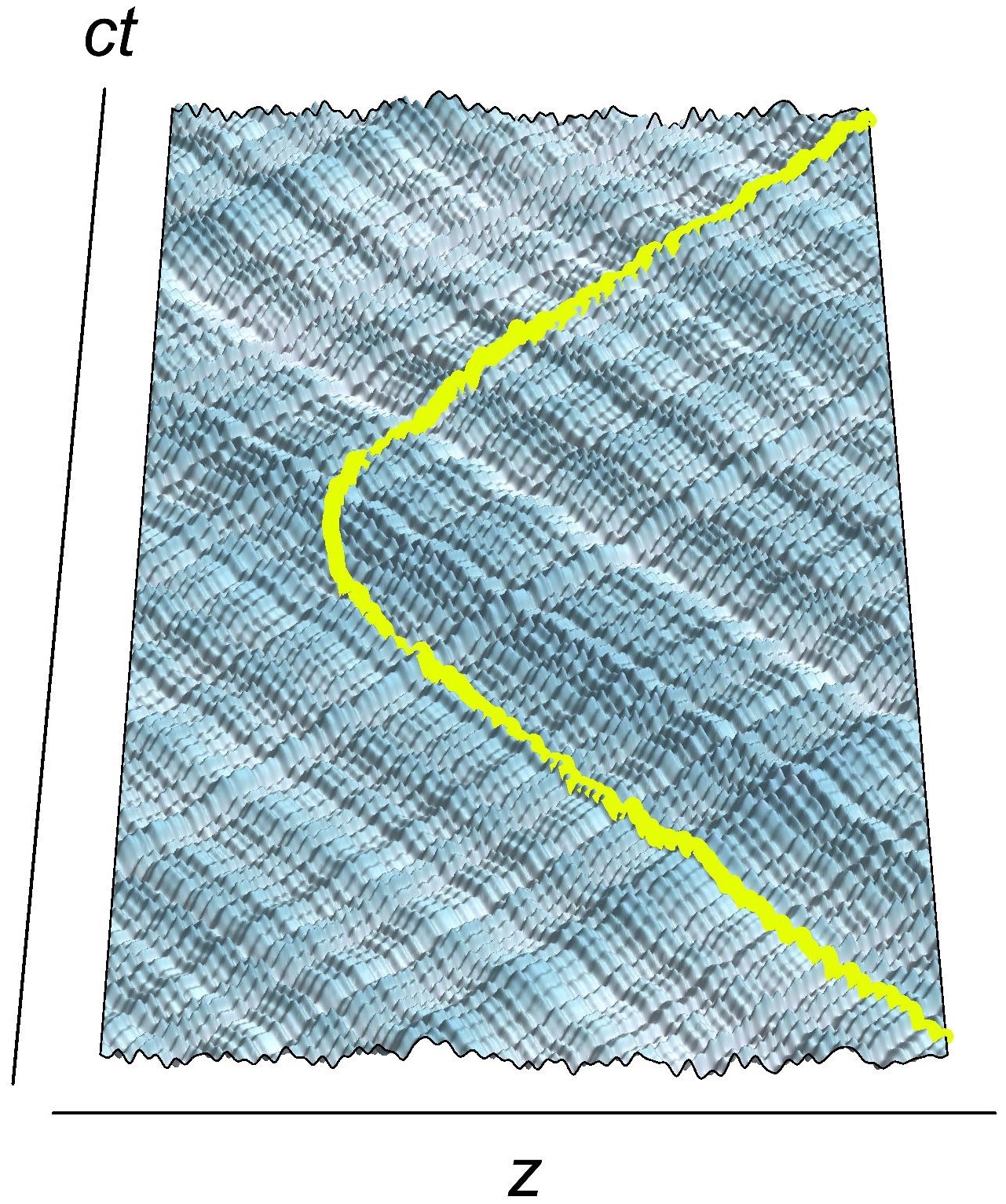}
\caption{
\small{
Principal idea. A container is filled with water subject to noise creating ripples on the water surface. Top: a movable mirror guides a laser beam over the water surface illuminating a sharp spot recorded by a video camera. Bottom: video of the water surface and space-time diagram of the illuminated spot following the trajectory of the accelerated observer (Fig.~\ref{diagram}). 
}
\label{scheme}}
\end{center}
\end{figure}

Here we propose and experimentally demonstrate a classical analogue of the Unruh effect, where $\hbar$ is replaced by the strength of classical noise and $c$ by the speed of the waves involved in the effect. In our case (Fig.~\ref{scheme}) these are water waves with $c$ of about $0.2\mathrm{m}/\mathrm{s}$. In this way, the Unruh temperature of Eq.~(\ref{unruh}) is boosted such that the Unruh effect becomes observable.

Analogues \cite{Analogues} of the Unruh effect have been proposed before: the use of impurities in Bose-Einstein condensates as accelerated particle detectors \cite{Retzker} or of graphene \cite{Graphene} folded into a Beltrami trumpet \cite{Needham} that corresponds to an accelerate space. It was also suggested \cite{Lewenstein} to employ a quantum simulator made of cold atoms in an optical lattice to generate a synthetic Unruh effect in arbitrary dimensions \cite{Takagi}. So far, none of these ideas, exciting as they are, were experimentally demonstrated. Connections between the Unruh effect and classical physics have also been pointed out before \cite{Boyer,Higuchi,Pauri}, but not the simple connection we found.

\begin{figure}[h]
\begin{center}
\includegraphics[width=15.0pc]{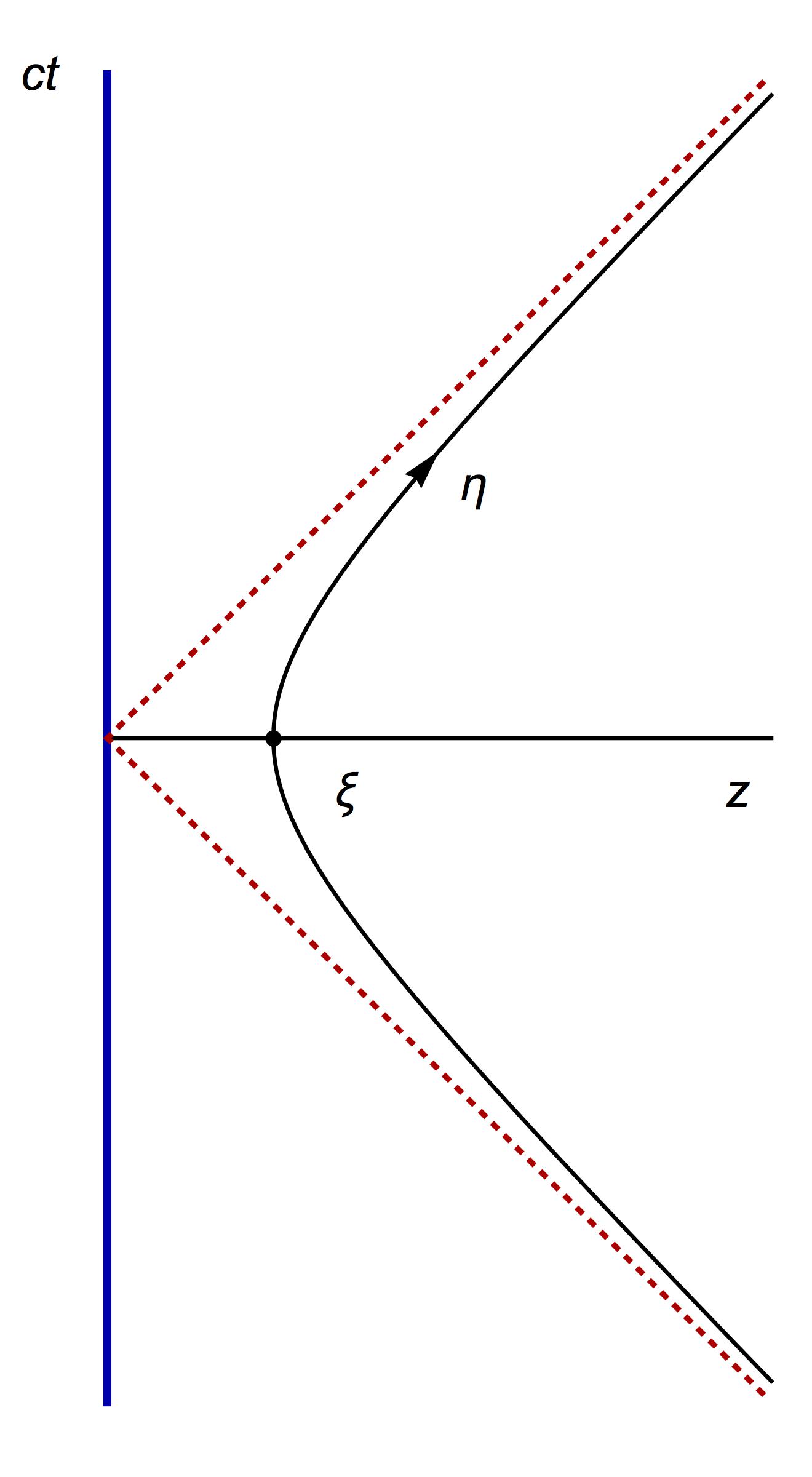}
\caption{
\small{
Space-time diagram. The accelerated observer follows a hyperbola (black curve) in space--time. The observer comes in from $\infty$ with asymptotically $-c$, gets slower due to the acceleration in positive direction until coming to rest for a fleeting moment at $z=\xi$, and changing direction. Then the observer gains speed, asymptotically approaching $+c$ at $\infty$. The red dotted lines indicate the causal cones straddled by the accelerated observer. The trajectory obeys Rindler's formula, Eq.~(\ref{rindler}), for constant $\xi$.
}
\label{diagram}}
\end{center}
\end{figure}

One advantage of our scheme is its simplicity. Figure \ref{scheme} illustrates the principal idea; the actual experiment is modified and described in Sec.~III. Imagine a container filled with water is subject to white noise. The resulting ripples on the water surface are scanned with a movable laser beam, while a camera is taking a video of the height of the illuminated spot \cite{Water,Germain}. The moving spot plays the role of the moving detector; the water ripples represent the vacuum noise. The spot should move such that its space--time trajectory matches the space--time diagram (Fig.~\ref{diagram}) of an observer with constant relativistic acceleration where $c$ is replaced by the speed of the water waves. The varying height of the water ripples are recorded along the trajectory for each run, and the experiment is repeated many times to get reliable statistics. 

Note that the combination of laser spot and video camera acts like an amplitude detector, whereas Unruh \cite{Unruh} considered a particle detector. However, an amplitude detector can, in principle, replace a particle detector: the particle-number distribution is tomographically obtainable from amplitude measurements \cite{Leo}. For example, in optical homodyne tomography \cite{Raymer} amplitude measurements are sufficient for reconstructing the quantum state of light \cite{Leo} that includes the results of photon detection. Here we do not use tomography, but develop a form of Fourier analysis where we directly read off the correlations in the Unruh effect that give the Planck spectrum.

These correlations are modified in an interesting way by the boundaries of the container. In free space, an accelerated observer gets quantum--entangled with a partner if such a partner moves on the exact mirror image of the observer's trajectory \cite{UnruhEntanglement,LeoBook}. Whenever the first observer records the click of a particle detector, so does the partner (assuming perfect detection efficiency). If the two paired observers use amplitude detectors, they record the two--mode squeezing \cite{LeoBook} of Gaussian noise. In our case (Fig.~\ref{scheme}) the boundary of the container acts like a mirror reflecting a hypothetical partner back onto the trajectory of the observer, which turns out to create single--mode squeezing of noise, an effect we have clearly observed experimentally.

Our findings suggest that  at the heart of the Unruh effect lies the correlation of wave noise, regardless whether these waves are quantum or classical. Figure \ref{scheme} (bottom) illustrates this idea. The figure shows the space--time diagram of water wave subject to noise. Although the waves amplitudes are random in space, they are organized in space--time: one clearly sees the causal cones of wave propagation, in addition to the reflections at the boundaries. This organization of wave noise in space and time generates the correlations in the Unruh effect that appear to a single observer as excess thermal energy with Unruh temperature, Eq.~(\ref{unruh}).

\section{Theory}

Let us begin with a miniature review on accelerated observers for introducing the notation and for keeping the paper as self--contained as possible. Figure \ref{diagram} shows the space--time diagram of the accelerated observer with position $z$ at time $t$; the detector of the observer follows a hyperbola parameterised in terms of the Rindler coordinates \cite{Rindler,LeoBook} $\xi$ and $\eta$ as
\begin{equation}
z = \xi \cosh \eta \,,\quad ct = \xi \sinh \eta
\label{rindler}
\end{equation}
with constant $\xi$. We briefly prove that the Rindler trajectory (\ref{rindler}) indeed describes constant acceleration \cite{Alsing}. For this, we express the Minkowski metric $\mathrm{d}s^2=c^2\mathrm{d}t^2-\mathrm{d}z^2$ in Rindler coordinates (\ref{rindler}) and get $\mathrm{d}s^2=\xi^2\mathrm{d}\eta^2-\mathrm{d}\xi^2$. The metric $s$ divided by $c$ gives the proper time $\tau$. Since $\mathrm{d}\xi=0$ for constant $\xi$ we obtain
\begin{equation}
\tau = \frac{\xi}{c}\,\eta \,.
\label{tau}
\end{equation}
The parameter $\eta$ is thus proportional to time $\tau$ as perceived by the accelerated observer. We get for the Rindler trajectory (\ref{rindler}) $\mathrm{d}z/\mathrm{d}\eta=ct$ and hence $\mathrm{d}z/\mathrm{d}\tau=(c^2/\xi)\,t$. From this follows for the relativistic acceleration (the force divided by the rest mass)
\begin{equation}
a = \frac{\mathrm{d}}{\mathrm{d}t}\,\frac{\mathrm{d}z}{\mathrm{d}\tau} = \frac{c^2}{\xi} \,,
\label{a}
\end{equation}
which is indeed a constant for constant $\xi$. The Rindler trajectory thus describes uniform acceleration.

\begin{figure}[h]
\begin{center}
\includegraphics[width=15.0pc]{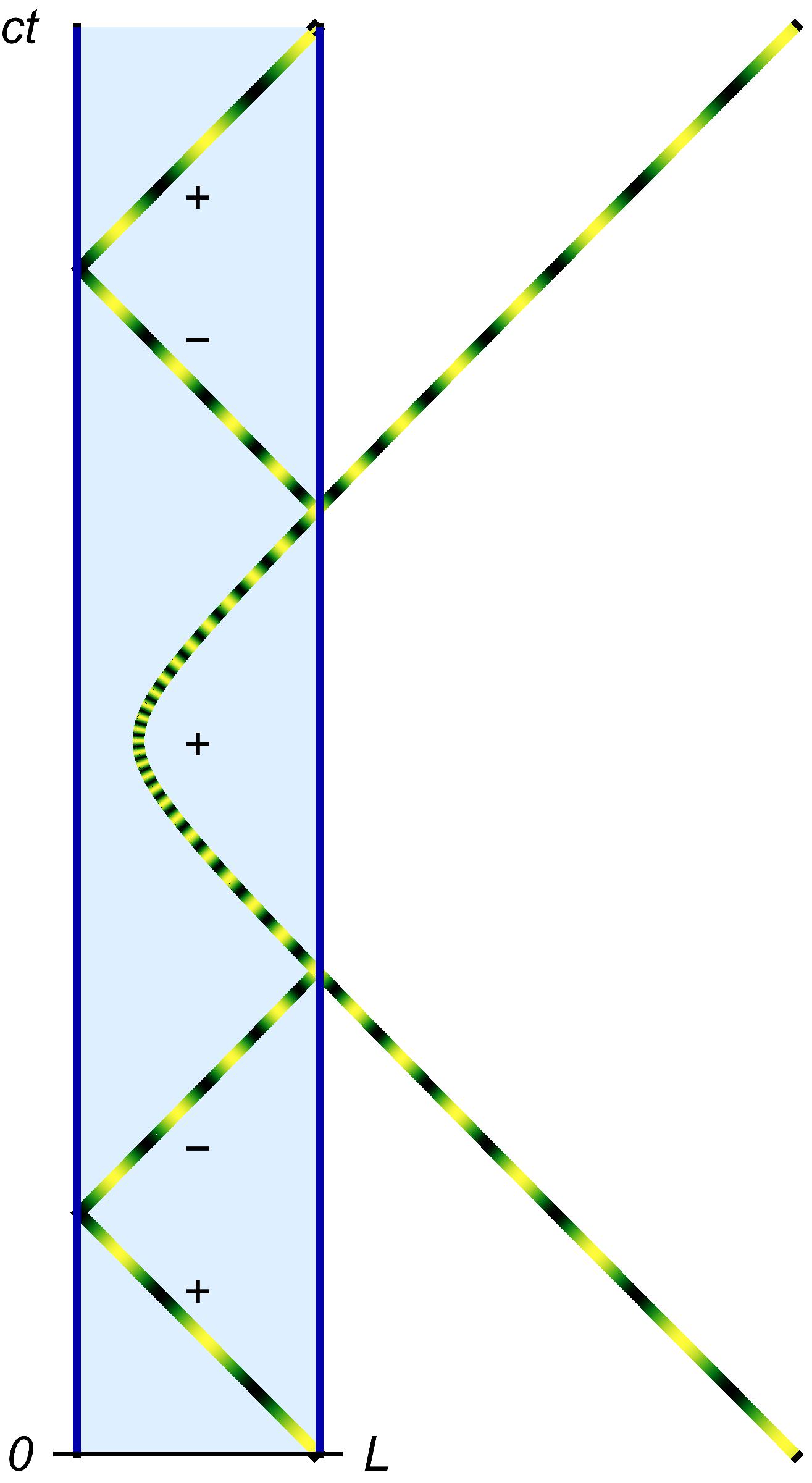}
\caption{
\small{
Mirrors. Two reflecting boundaries or nodes, acting as mirrors, confine the waves between $0$ and $L$. In this case, instead of tracing the full Rindler trajectory (Fig.~\ref{diagram}) it is sufficient to scan its mirror images with the appropriate signs indicated. The pulsation along the space--time trajectory indicates the changing measure of time experienced by the accelerated observer. As time flows exponentially slowly for velocities approaching $c$, an exponentially large lab would be required to trace a sufficiently long trajectory, were it not for the mirrors.}
\label{mirrors}}
\end{center}
\end{figure}

In our experiment, $c$ is replaced by the speed of the water waves; the amplitude detector should follow the corresponding Rindler trajectory of Eq.~(\ref{rindler}). We made another simplification that makes the experiment feasible: the water channel cannot be infinitely extended, but shall have reflecting boundaries or nodes that act as mirrors for water waves (Fig.~\ref{mirrors}). The mirror on the left is placed at the origin ($z=0$) of the Rindler frame --- at the origin of the causal cone the accelerated observer straddles; the mirror at the right ($z=L$) is less important in principle, but very important in practice: as the two mirrors reflect the waves, one does not need to trace the entire Rindler trajectory, but only its reflections in the mirrors (Fig.~\ref{mirrors}). Since $z$ grows exponentially with $\eta$, the pair of mirrors saves exponentially large lab space.

The amplitude $A$ of the water surface can be understood as a superposition of modes $A_k$ with coefficients $\alpha_k$ (if the wave propagation is linear):
\begin{equation}
A = \int_0^\infty \big(\alpha_k A_k + \alpha_k^* A_k^*\big) \,\mathrm{d}k \,.
\label{expansion}
\end{equation}
The mode coefficients $\alpha_k$  encode the physical state of the wave, including their noise. The coefficients are complex numbers written in terms of the quadratures $q$ and $p$ \cite{LeoBook} as
\begin{equation}
\alpha = \frac{1}{\sqrt{2}}(q+\mathrm{i}p) \,.
\label{quadratures}
\end{equation}
We assume Gaussian noise of uniform strength $I$ for the quadratures such that the averages $\langle q\rangle$, $\langle p\rangle$ and $\langle q p \rangle$ vanish, and 
\begin{equation}
\langle q_1 q_2 \rangle = \langle p_1 p_2 \rangle = \frac{I}{2}\,\delta(k_1-k_2) \,.
\label{noise}
\end{equation}
For defining the strength of the noise we need to normalise the modes according to a certain time--invariant scale. For this we use the scalar product
\begin{equation}
\left(A_1,A_2\right) = \frac{\mathrm{i}}{c}\int \left( A_1^* \frac{\partial A_2}{\partial t} - A_2 \frac{\partial A_1^*}{\partial t}\right)\,\mathrm{d}z
\label{scalar}
\end{equation}
that is invariant in time for modes satisfying the wave equation. The left mirror enforces the boundary condition $A_k=0$ at $z=0$ and thus selects from the plane waves with wavenumbers $k$ the superposition
\begin{equation}
A_k = {\cal A}\sin (kz) \exp(-\mathrm{i} kct) \,.
\label{modes}
\end{equation}
These modes are normalized to $\delta(k_1-k_2)$ according to the scalar product of Eq.~(\ref{scalar}) for 
\begin{equation}
{\cal A} = \frac{1}{\sqrt{\pi k}} \,.
\label{amplitude}
\end{equation}
The right mirror at $z=L$ imposes 
\begin{equation}
k = m \frac{\pi}{L} \,.
\end{equation}
With this set of wavenumbers the amplitude $A$ would, mathematically, be a periodic function in space, $A(z+2L)=A(z)$, and, as $A(-z)=-A(z)$, we have $A(z+L)=-A(L-z)$. This means that instead of scanning the entire trajectory of the accelerated observer, we only need to scan its reflections with the appropriate signs (Fig.~\ref{mirrors}). 

In the following we ignore the auxiliary right mirror (assuming a sufficiently dense set of modes). Suppose that a statistical ensemble of many videos of the waves are taken. In the original Unruh effect \cite{Unruh}, a Planck spectrum with the temperature of Eq.~(\ref{unruh}) is predicted for the accelerated observer. In order to get information about the spectrum, we need to Fourier transform the recorded wave amplitudes along the Rindler trajectories of Eq.~(\ref{rindler}) and for the proper time as seen by the accelerated observer, Eq.~(\ref{tau}), {\it i.e.} with respect to $\eta$:
\begin{equation}
\widetilde{A} = \int_{-\infty}^{+\infty} A \,\mathrm{e}^{\mathrm{i}\nu\eta} \,\mathrm{d}\eta \,.
\label{fourier}
\end{equation}
This is the experimental quantity of interest we need to analyse and compare with the Unruh effect \cite{Unruh,Fulling,Davies}.

\begin{figure}[t]
\begin{center}
\includegraphics[width=15.0pc]{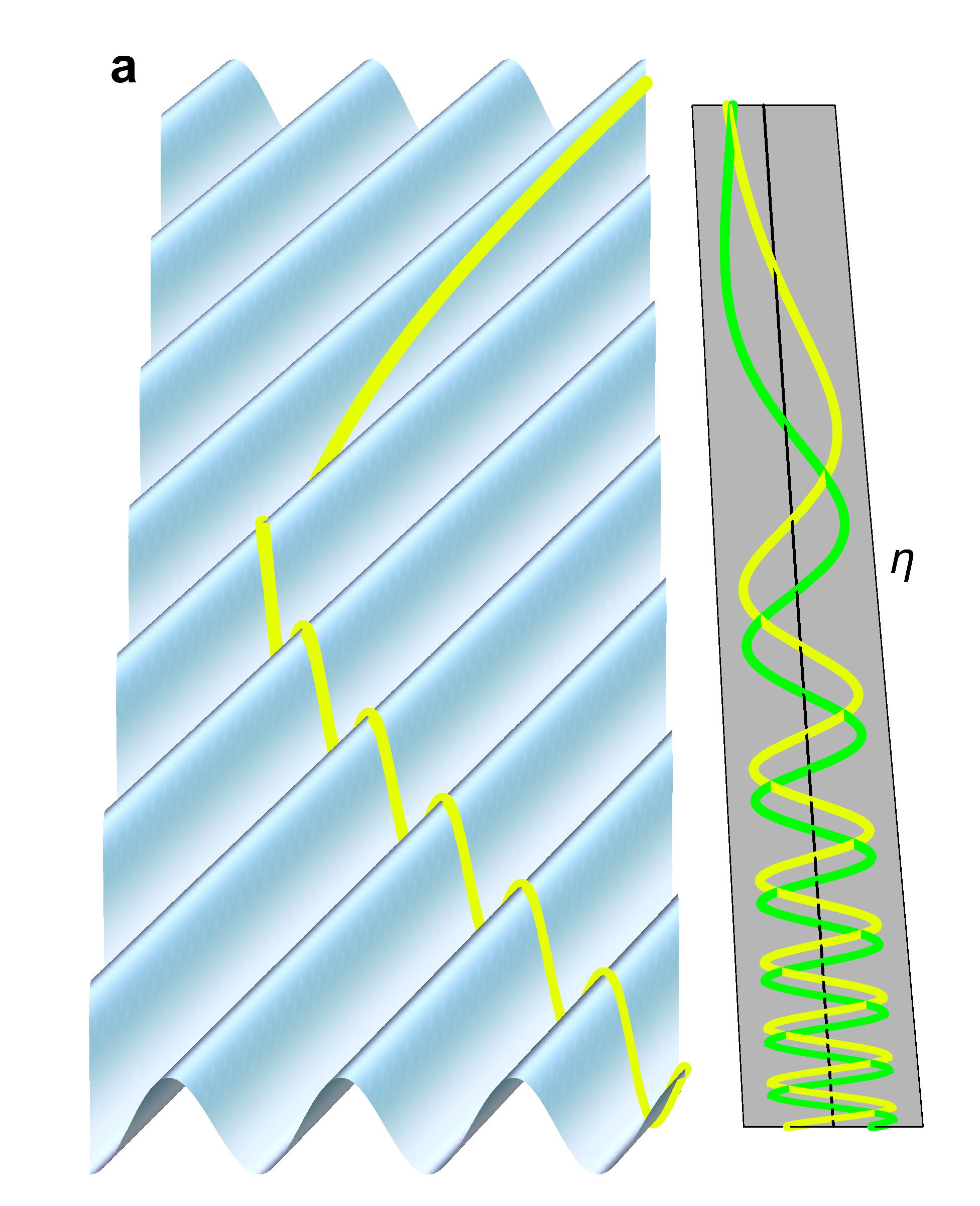}
\includegraphics[width=18.0pc]{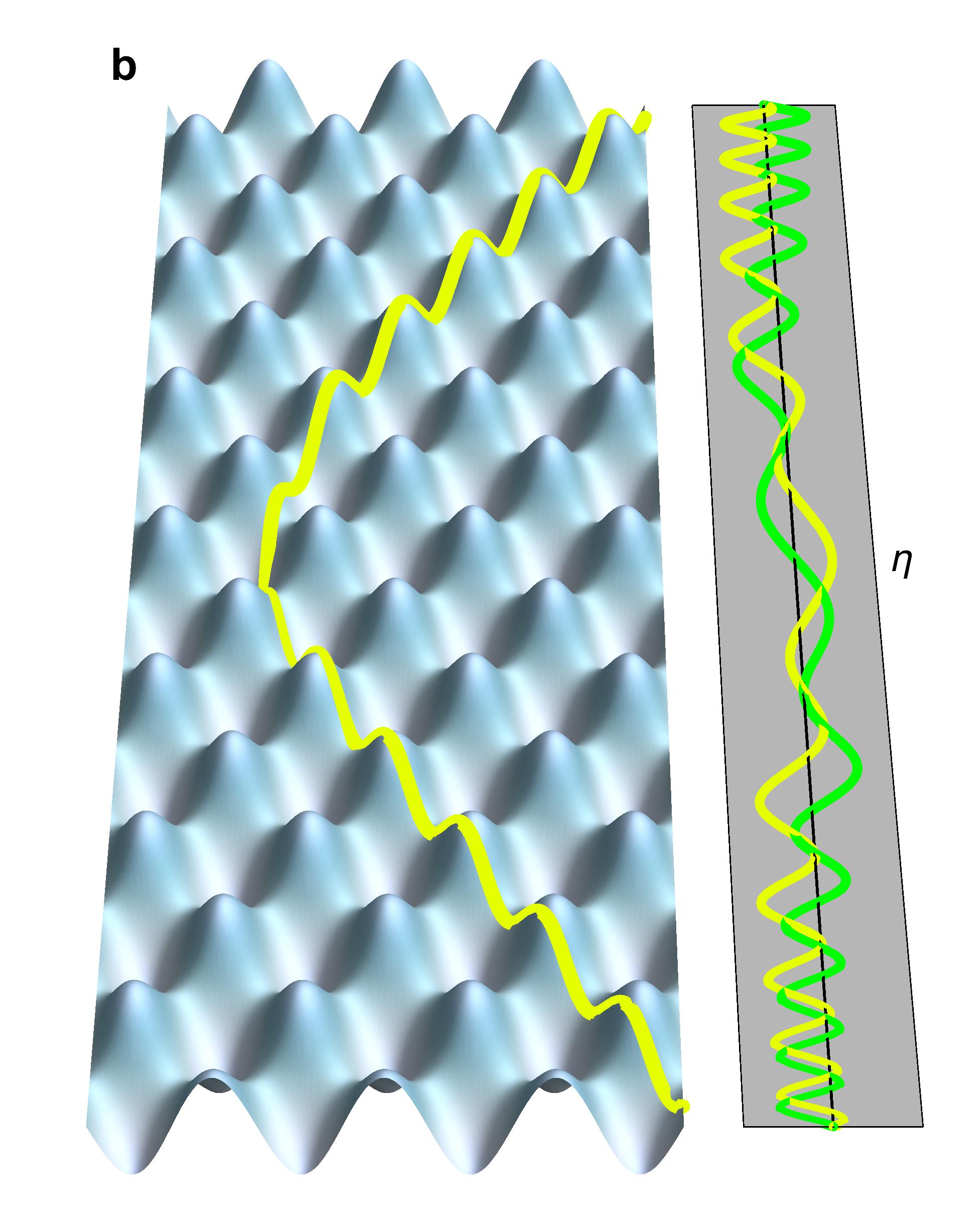}
\caption{
\small{
Plane waves. {\bf a}: the accelerated observer (Fig.~\ref{diagram}) traces a single running plane wave; on the side panel the real part (yellow) and imaginary part (green) of the signal are plotted as functions of $\eta$. One sees exponentially rapid oscillations for $\eta\ll-1$ and an exponential freeze for $\eta\gg 1$. {\bf b}: the observer traces a standing wave. The real part (yellow) is an even function in $\eta$, the imaginary part (green) is odd in $\eta$, both oscillate exponentially for $|\eta|\gg 1$.
}
\label{trace}}
\end{center}
\end{figure}

As the amplitude $A$ is the superposition of modes $A_k$ according to Eq.~(\ref{expansion}), we focus on one arbitrary mode, Eq.~(\ref{modes}), and express it in the Rindler coordinates of Eq.~(\ref{rindler}):
\begin{equation}
A_k = \frac{{\cal A}}{2\mathrm{i}}\left[\exp\left(\mathrm{i}k\xi\mathrm{e}^{-\eta}\right)-\exp\left(-\mathrm{i}k\xi\mathrm{e}^{\eta}\right)\right] \,.
\label{rindlermodes}
\end{equation}
Consider either of the two plane waves that constitute $A_k$ (Fig.~\ref{trace}a) \cite{Alsing}. We obtain for the Fourier transform 
\begin{eqnarray}
\widetilde{A_\pm} & = & \int_{-\infty}^{+\infty} \exp\big(\pm\mathrm{i} k\xi \mathrm{e}^{\mp \eta} + \mathrm{i}\nu\eta\big) \mathrm{d}\eta
\label{f1} \\
& = & \mp \left(\mp\mathrm{i} k\xi\right)^{\pm\mathrm{i}\nu} \int_0^{\pm\mathrm{i}\infty} \mathrm{e}^{-x} x^{\mp\mathrm{i}\nu-1} \mathrm{d} x
\nonumber \\
& = & - (k\xi)^{\pm\mathrm{i}\nu} \mathrm{e}^{\pi\nu/2} \,\Gamma(\mp\mathrm{i}\nu)
\label{result0}
\end{eqnarray}
where we substituted $x=\mp\mathrm{i}k\xi\,\mathrm{e}^{\mp \eta}$ in the first step and deformed the integration contour to the real axis in the second step, using there also the definition of the gamma function \cite{Erdelyi} and $(\mp\mathrm{i})^{\pm\mathrm{i}\nu}=\mathrm{e}^{\pi\nu/2}$. Now, turn to the Fourier integral of the complex conjugate plane wave:
\begin{equation}
\widetilde{A_\pm^*} = \int_{-\infty}^{+\infty} \exp\big(\mp\mathrm{i} k\xi \mathrm{e}^{\mp \eta} + \mathrm{i}\nu\eta\big) \mathrm{d}\eta \,.
\label{f2}
\end{equation}
Substituting $x=\pm\mathrm{i}k\xi\,\mathrm{e}^{\mp \eta}$ and using $(\pm\mathrm{i})^{\pm\mathrm{i}\nu}=\mathrm{e}^{-\pi\nu/2}$ in this case, one obtains the remarkable relation \cite{Remark}
\begin{equation}
\widetilde{A_\pm^*} = \mathrm{e}^{-\pi\nu} \widetilde{A_\pm} \,.
\label{exp}
\end{equation}
The factor $\mathrm{e}^{-\pi\nu}$ is exponential in $\nu$ and independent of the mode index, which turns out to be the mathematical key to the thermality and universality of the Unruh effect.

Having obtained the results of Eqs~(\ref{result0}) and (\ref{exp}) for running plane waves \cite{Alsing}, we turn to the standing waves of Eq.~(\ref{rindlermodes}) --- our modes (Fig.~\ref{trace}b). We get for their Fourier transforms 
\begin{eqnarray}
\widetilde{A_k} &=& - {\cal A}\, \mathrm{e}^{\pi\nu/2} \mathrm{Im}\left[ (k\xi)^{\mathrm{i}\nu}\Gamma(-\mathrm{i}\nu)\right]
\nonumber\\
&=& - \mathrm{e}^{\pi\nu/2} \,\frac{\sin(\nu\ln k\xi - \phi)}{\sqrt{k\nu\sinh\nu\pi}} \,.
\label{aktilde}
\end{eqnarray}
In the last step we have used Eq.~(\ref{amplitude}) for ${\cal A}$ and the relationship $|\Gamma(\mathrm{i}\nu)|^2=\pi/(\nu \sinh\pi\nu)$ for the magnitude of the gamma function \cite{Erdelyi}; $\phi$ abbreviates the phase $\arg\Gamma(\mathrm{i}\nu)$ \cite{Phase}. For the Fourier transforms of the complex conjugate modes we have as before:
\begin{equation}
\widetilde{A_k^*} = \mathrm{e}^{-\pi\nu} \widetilde{A_k} \,.
\label{akctilde}
\end{equation}
We substitute Eqs.~(\ref{aktilde}) and (\ref{akctilde}) into the mode expansion, Eq.~(\ref{expansion}), of the Fourier integral, Eq.~(\ref{fourier}), and arrive at the expression
\begin{equation}
\widetilde{A} = \int_0^\infty \frac{\sin(\phi-\nu\ln k\xi)}{\sqrt{k\nu\sinh\nu\pi}}\,\big(\alpha_k \mathrm{e}^{\pi\nu/2} + \alpha_k^* \mathrm{e}^{-\pi\nu/2} \big) \,\mathrm{d}k\,.
\label{result}
\end{equation}
It is wise to combine the $\alpha_k$ in Eq.~(\ref{result}) in the total amplitude
\begin{equation}
\alpha = \int_0^\infty  \frac{\sin(\phi-\nu\ln k\xi)}{\sqrt{\pi k}} \,\alpha_k\, \mathrm{d}k \,.
\label{total}
\end{equation}
Given that the individual mode amplitudes $\alpha_k$ represent Gaussian noise, the total amplitude $\alpha$ is Gaussian as well. Given the only non--vanishing second moments of Eq.~(\ref{noise}) for the individual quadratures, the quadratures of the total amplitude must fluctuate with the same strength \cite{Moments}:
\begin{equation}
\langle q(\nu_1) q(\nu_2) \rangle = \langle p(\nu_1) p(\nu_2) \rangle =\frac{I}{2}\,\delta(\nu_1-\nu_2) \,.
\label{numoments}
\end{equation}
Gaussian noise is completely characterised by the first and second moments, so the total mode amplitude $\alpha$ represents exactly the same noise as each of the individual mode amplitudes. 

The amplitude $\alpha$ describes the total noise incident in one Fourier component of the detected signal, the total incident noise, but this is not the noise detected by the moving observer. To determine the detected noise we represent the exponential factor $\mathrm{e}^{-\pi\nu}$ as
\begin{equation}
\mathrm{e}^{-\pi\nu} = \tanh\zeta \,.
\label{tanh}
\end{equation}
We express the Fourier transformed amplitude along the Rindler trajectory, Eq.~(\ref{result}), in terms of the total noise amplitude, Eq.~(\ref{total}), and its quadratures, Eq.~(\ref{quadratures}), and arrive at the compact expressions
\begin{eqnarray}
\widetilde{A} &=& \sqrt{\frac{2}{\nu}}\,\big(\alpha\cosh\zeta+\alpha^*\sinh\zeta\big) 
\nonumber\\
&=& \frac{1}{\sqrt{\nu}}\big(q\,\mathrm{e}^\zeta + \mathrm{i}p\,\mathrm{e}^{-\zeta}\big) \,.
\end{eqnarray}
We see that the detected noise is squeezed --- the noise in the $p$ quadrature is reduced at the expense of the noise in the $q$ quadrature \cite{LeoBook}. The squeezing parameter $\Delta (\mathrm{Re}\widetilde{A})/\Delta (\mathrm{Im}\widetilde{A})=\mathrm{e}^{2\zeta}$ we easily obtain solving Eq.~(\ref{tanh}) for $\mathrm{e}^{2\zeta}$:
\begin{equation}
\frac{\Delta (\mathrm{Re}\widetilde{A})}{\Delta (\mathrm{Im}\widetilde{A})}= \coth \frac{\pi\nu}{2} \,.
\label{squeezing}
\end{equation}
Note that although the detected noise is reduced in $\mathrm{Im}\widetilde{A}$, the total noise has grown:
\begin{eqnarray}
\langle \widetilde{A}(\nu_1)\widetilde{A}^*(\nu_2)\rangle
&=&\frac{I}{2\nu}\left(\mathrm{e}^{2\zeta} + \mathrm{e}^{-2\zeta}\right) \delta(\nu_1-\nu_2)
\nonumber\\
&=& \frac{2}{\nu}\,I \left(\frac{1}{2} + \sinh^2\zeta\right) \delta(\nu_1-\nu_2) \,.
\end{eqnarray}
Here the $1/2$ represents the incident noise --- the equivalent of the vacuum noise, while the $\sinh^2\zeta$ term accounts for the additional fluctuations perceived in total by the moving observer. We denote $\sinh^2\zeta$ by $\overline{n}$ and obtain from Eq.~(\ref{tanh}):
\begin{equation}
\overline{n} = \frac{1}{\mathrm{e}^{2\pi\nu}-1} \,.
\label{sinh2}
\end{equation}
The Fourier component $\nu$ to the dimensionless Rindler parameter $\eta$ is proportional to the frequency $\omega$ with respect to the proper time $\tau$ of the moving observer. We get from Eqs.~(\ref{tau}) and (\ref{a}):
\begin{equation}
\nu = \frac{\xi}{c}\,\omega = \frac{c}{a}\,\omega \,.
\label{relation}
\end{equation}
Reading $2\pi\nu$ in Eq.~(\ref{sinh2}) as $\hbar\omega/K_\mathrm{B}T$ we see that the energy of the extra noise $\overline{n}$ follows a Planck distribution; using Eq.~(\ref{relation}) we realise that its temperature $T$ matches exactly the Unruh temperature of Eq.~(\ref{unruh}).

Our water--wave analogue exactly reproduces the Unruh effect for the total fluctuations; the squeezing is due to the mirror. Without the mirror the signal along the Rindler trajectory would be correlated to the signal along the mirror image of the trajectory. The mirror projects these correlations into the Fourier quadratures of a single trajectory; two--mode squeezing \cite{LeoBook} of noise turns into single--mode squeezing \cite{LeoBook}. Our analogue shows the essence of the correlations in the Unruh effect \cite{UnruhEntanglement} with an interesting twist.

\section{Experiment}

We performed an experiment to test whether these ideas are robust under real laboratory conditions. For this, we simplified our scheme (Fig.~\ref{scheme}) even further. Instead of taking the video of the height of the water surface at a moving spot representing the accelerated observer on a Rindler trajectory (Fig.~\ref{diagram}), we took a video of the entire surface evolving in time. We then analyzed {\it a--posteriori} the measured surface along Rindler trajectories, described by Eq.~(\ref{rindler}), varying $\xi$ and hence, according to Eq.~(\ref{a}), the acceleration $a$.

We also did not apply white noise to the water, but rather created a standing wave through Faraday instability \cite{Faraday} by oscillating vertically the container. Such Faraday waves behaves like laser light --- they have stable average amplitudes due to the balance of gain and loss, but carry some amplitude noise. We randomized the phase for having a complete analogue to laser light. With this, we studied the {\it stimulated} Unruh effect similar to the experiments \cite{Water} on the stimulated Hawking effect in water. The stimulated effect shares the characteristic features of the Unruh effect --- the quadrature squeezing according to the Planck spectrum with the correct temperature, Eq.~(\ref{unruh}). This type of experiment has the advantage of avoiding dispersion --- the wavelength dependance of $c$, because only one wavelength is used. Without dispersion, $c$ is always well--defined and can therefore be used without restriction as the basis for the Rindler trajectories of Eq.~(\ref{rindler}).

\begin{figure}[h]
\begin{center}
\includegraphics[width=20pc]{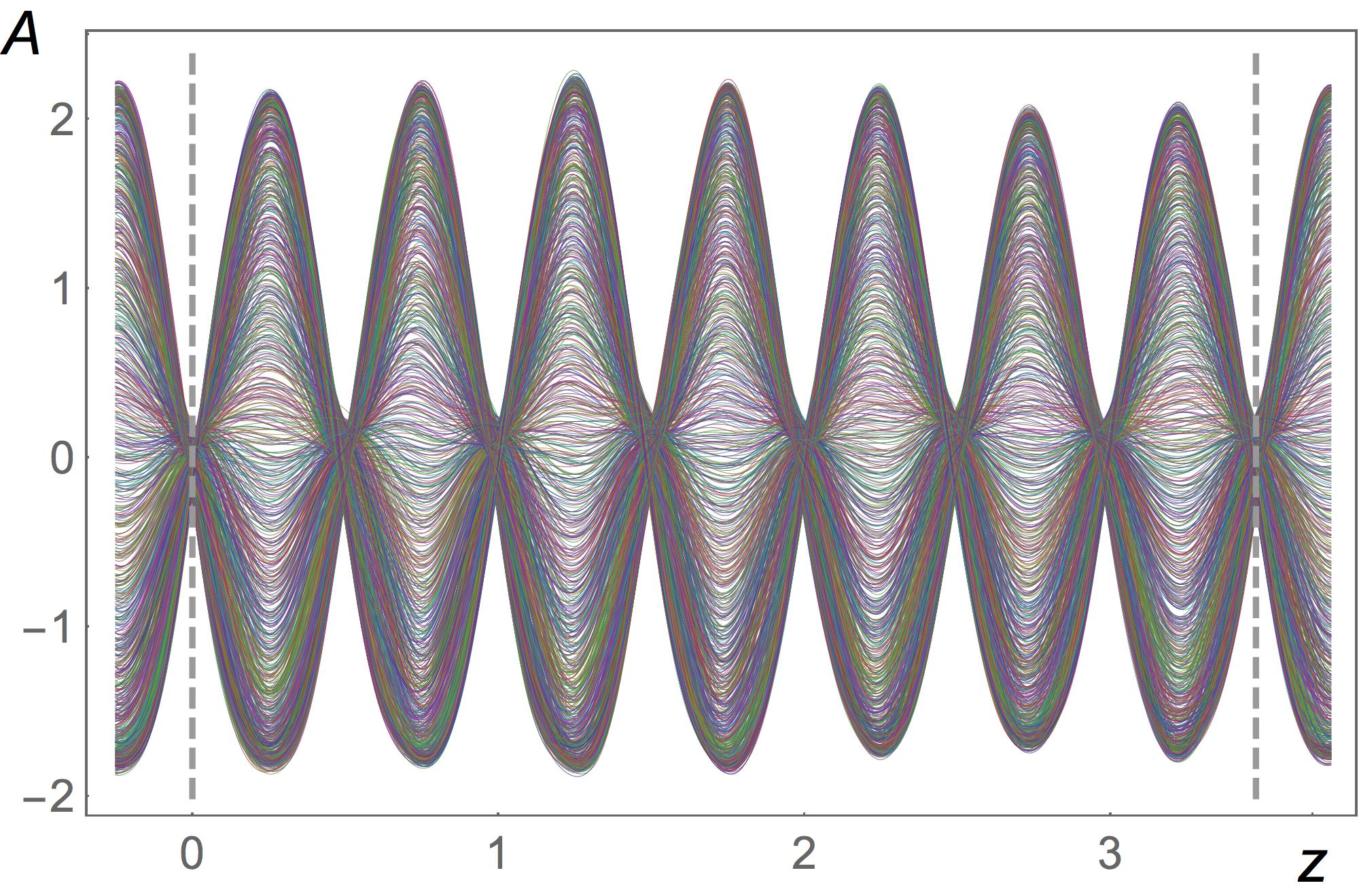}
\includegraphics[width=20pc]{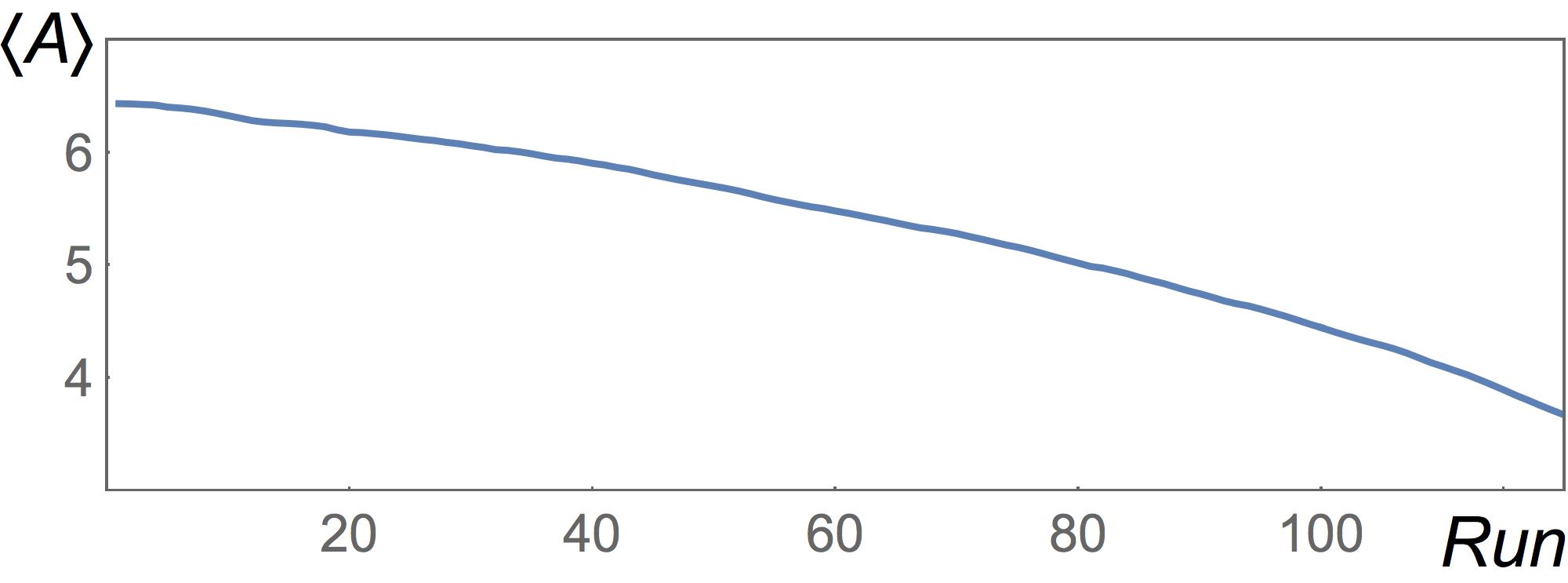}
\caption{
\small{
Waves produced through Faraday instability. Top: measured wave amplitudes $A$ in arbitrary units along $z$ in units of wavelength for the first $100$ cycles of wave propagation. The wave pattern continues to the left and right of the figure, but with decreasing amplitude. We selected two nodes of the standing waves as our mirrors (dashed lines). One sees that the waves are not perfectly harmonic --- Fourier analysis (not shown here) reveals that anharmonicities contribute to about $10\%$ of the amplitude. Bottom: decline of the amplitude averaged over each cycle $\langle A\rangle$ as a function of runs where we divided the data into pieces of hundred cycles each with randomized phase. We corrected for this decline in our data analysis. 
}
\label{standingwaves}
}
\end{center}
\end{figure}

The experimental details are as follows. A standing wave field was created by exciting the Faraday instability \cite{Faraday} on the surface of a bath of plain tap water. The bath was vertically oscillated at a frequency of $19\mathrm{Hz}$ with an amplitude just above the instability threshold, giving rise to waves with a frequency of $9.5\mathrm{Hz}$ and a wavelength of $24\mathrm{mm}$. The rectangular shape of the water cavity ($250\mathrm{mm} \times 55 \mathrm{mm}$) ensured that an approximately one--dimensional standing wave formed along the length of the container. The profile of the water surface was measured by tracking the optical distortions of a striped floor pattern (seen through the liquid) using a digital video camera (at $500$ frames per second) and basic image processing. The resulting displacement field is proportional to the local slope of the water surface, which was numerically integrated to yield the height field \cite{Height}. The integration constant for each frame was determined from the conservation of mass. Data was taken for $1400\mathrm{s}$. Figure \ref{standingwaves} shows the standing--wave pattern for the first $100$ cycles. The figure also shows the gradual decline of the amplitude averaged over one cycle over time due to slow variations of the Faraday instability threshold; we corrected for this systematic decline in our data analysis. 

\begin{figure}[h]
\begin{center}
\includegraphics[width=16.8pc]{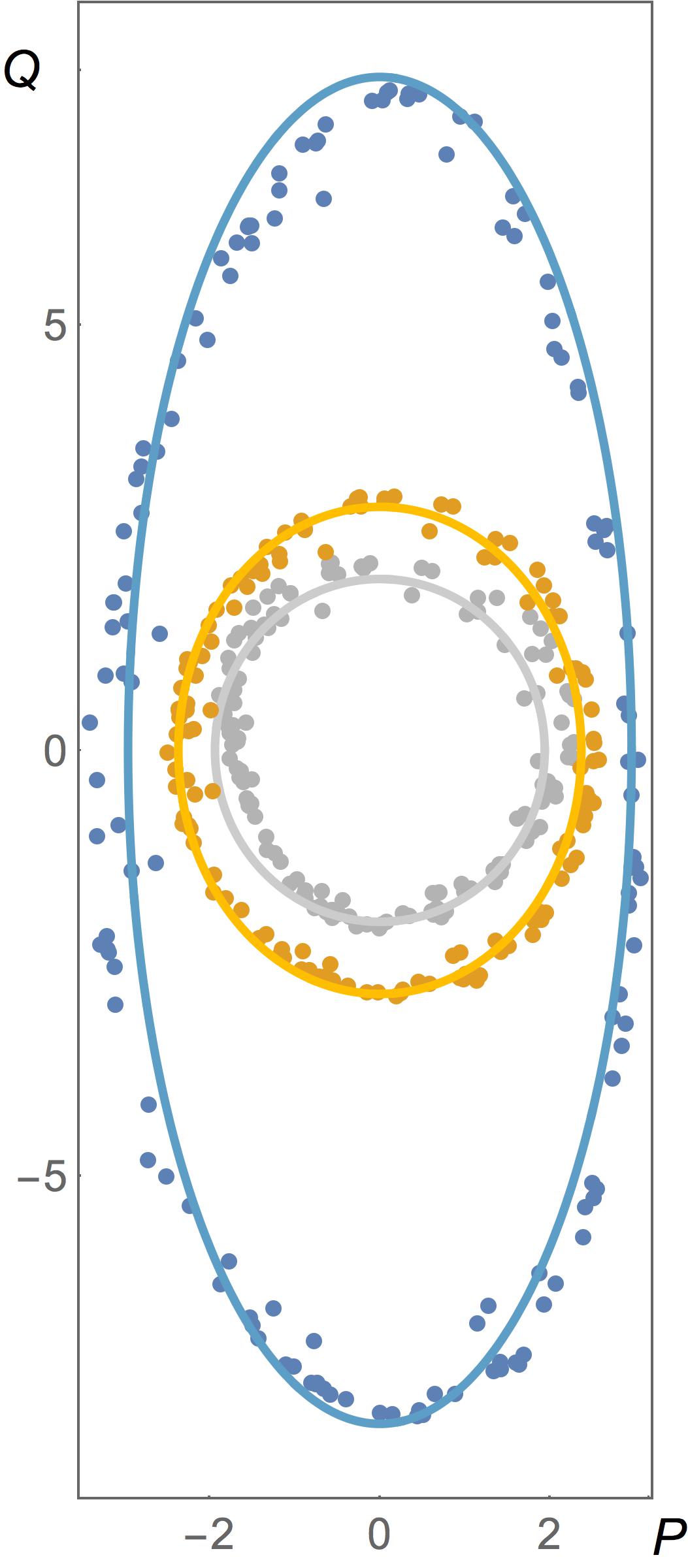}
\caption{
\small{
Experimental results. The dots show the real and imaginary half--odd Fourier coefficients in the units of $A$ (Fig.~\ref{standingwaves}) for each run of the experimental data,  blue: $\nu=1/4$, orange: $\nu=3/4$, gray: $\nu=5/4$; $Q=\mathrm{Re}\widetilde{A}$, $P=\mathrm{Im}\widetilde{A}$. The Fourier coefficients are taken according to Eq.~(\ref{Fcoeff}) along the space--time trajectory of an accelerated observer (Fig.~\ref{diagram}). The ellipses represent the theory, assuming the squeezing of noise with fixed amplitude (matched with the data) and random phase; the squeezing parameter is given by Eq.~(\ref{squeezing}).
}
\label{results}}
\end{center}
\end{figure}

Figure.~\ref{results} shows the results of the data analysis obtained with the method described in Appendix A: half--odd Fourier transformation. We selected from the $1400\mathrm{s}$ of data $131$ disjoined runs with $100$ cycles each, choosing a random initial phase for each run, and correcting for the systematic decline in average amplitude (bottom of Fig.~\ref{standingwaves}). Each run represents an individual element of a statistical ensemble with random phase (and with some amplitude noise). We choose a Rindler trajectory (Fig.~\ref{diagram}) with fixed parameter $\xi$ according to Eq.~(\ref{rindler}) and $\nu$ running from $-2\pi$ to $+2\pi$. When necessary, we mirrored the space--time trajectory (Fig.~\ref{mirrors}). Having choosen the trajectory, we calculated, for each run, the Fourier coefficients
\begin{equation}
\widetilde{A} = \int_{-2\pi}^{+2\pi} A \,\mathrm{e}^{\mathrm{i}\nu\eta} \,\mathrm{d}\eta 
\label{Fcoeff}
\end{equation}
for the first three half--odd Fourier numbers $\nu$ according to Eq.~(\ref{nu}): $\nu\in\{1/4,3/4,5/4\}$. Figure \ref{results} displays the real and imaginary part of the half--odd Fourier coefficients and compares them with theory --- the squeezed noise of a wave with fixed amplitude and random phase, with squeezing parameter given by Eq.~(\ref{squeezing}). One sees that the experiment agrees reasonably well with theory for the first two Fourier coefficients, despite the imperfections of the experiment, in particular the anharmonic contributions to the waves (Fig.~\ref{standingwaves}). 

\begin{figure}[h]
\begin{center}
\includegraphics[width=20pc]{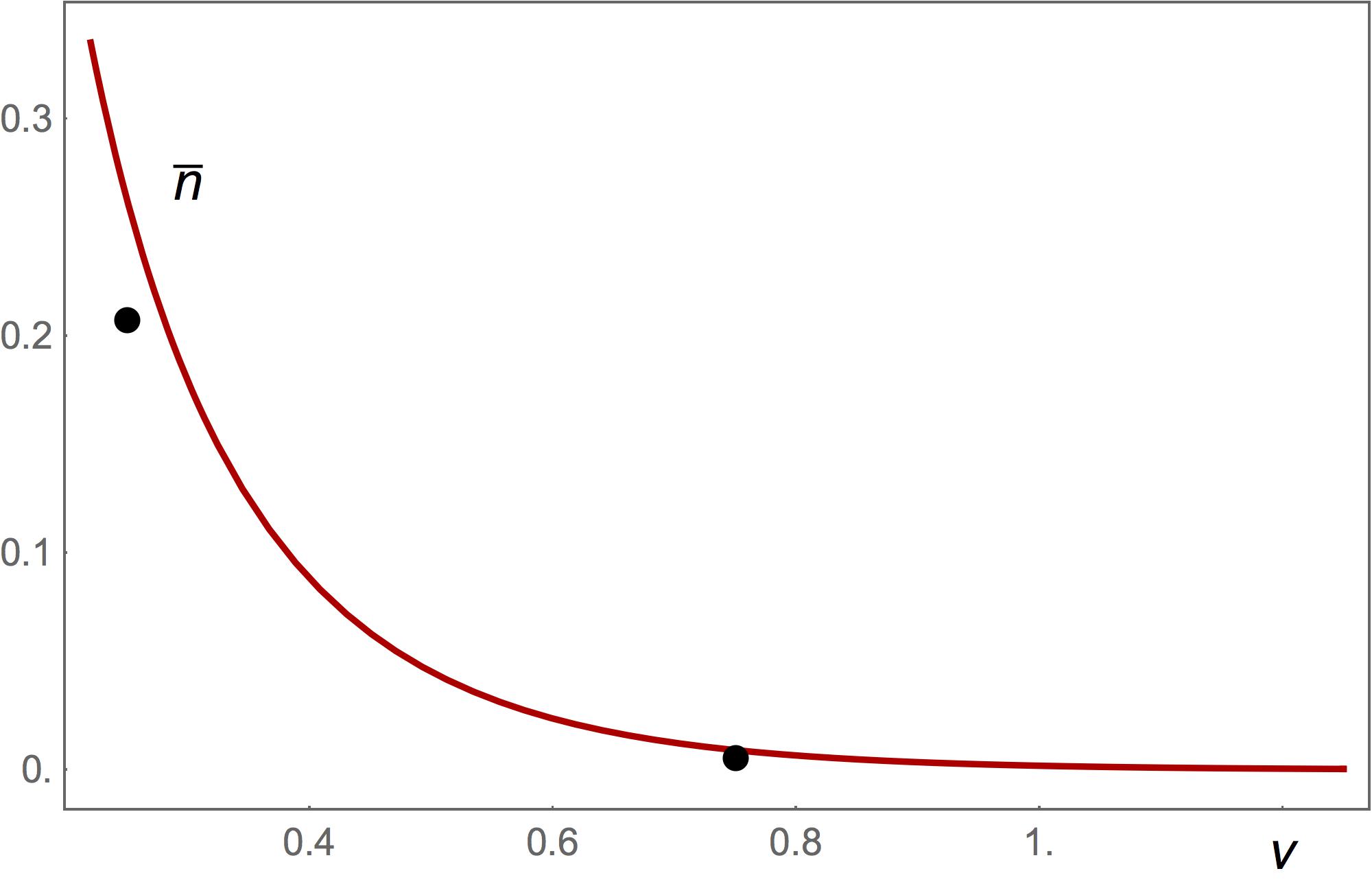}
\caption{
\small{
Planck curve. Dots: squeezing energy/ excess noise $\overline{n}$ calculated from the data (Fig.~\ref{results}) for $\nu=1/4$ and $\nu=3/4$. To obtain the two dots shown here, centered ellipses are fitted to the two corresponding data sets of Fig.~\ref{results}. From the ellipses the squeezing energy is calculated. Curve: theoretical prediction of a Planck curve according to Eq.~(\ref{sinh2}). The experimental points lie remarkably close to the theoretical curve, despite clear deviations of the waves from harmonicity (Fig.~\ref{standingwaves}), which illustrates the robustness of the Unruh effect against experimental imperfections. 
}
\label{planckcurve}}
\end{center}
\end{figure}

To quantify the squeezing energy, we calculated $\overline{n}$ as follows. We fitted centered ellipses to the data points of Fig.~\ref{results} by fitting a linear function  $Q^2 = \Delta Q^2 - (\Delta Q/\Delta P)^2 P^2$ to the points, with $Q=\mathrm{Re}\widetilde{A}$ and $P=\mathrm{Im}\widetilde{A}$. The linear coefficient of the fit directly gives $\Delta Q/\Delta P = \mathrm{e}^\zeta$, from which one obtains $\overline{n}=\sinh^2\zeta$. Our results are shown in Fig.~\ref{planckcurve} and compared with the Planck curve of Eq.~(\ref{sinh2}). 

From the statistical errors of the coefficients of the linear fit we determined the statistical errors of $(\Delta Q/\Delta P)^2$. We get $0.15$ for $\nu=1/4$ and $0.03$ for $\nu=3/4$. These errors are too small to explain the difference between the experimental values, $5.83$ and $1.34$, and the theoretical ones, $7.16$ and $1.46$, which shows that there are systematic errors in the data, most probably due to anharmonicities (Fig.~\ref{standingwaves}). Nevertheless, the agreement with theory in the squeezing ellipses (Fig.~\ref{results}) and in the Planck curve (Fig.~\ref{planckcurve}) is still remarkable. 

We varied $\xi$ and did not see much principal variation in the results, except that the agreement with theory gets better the larger $\xi$ is --- the smaller the acceleration $a$ is --- according to Eq.~(\ref{a}). The reason is probably the following: for smaller $a$ the space--time trajectory spends more proper time away from the node at $z=0$ where contributions from anharmonicity and other noise matter most. Figure~\ref{results} shows our results for the maximal $\xi$ we can accommodate for $-2\pi\le\nu\le+2\pi$ within $100$ cycles of wave oscillations. 

The third Fourier coefficient reveals the limits of the present experiment; there the subtle squeezing described by $\Delta (\mathrm{Re}\widetilde{A})/\Delta (\mathrm{Im}\widetilde{A})=\mathrm{coth}(\pi\nu/2)\approx 1.04$ for $\nu=5/4$ can no longer be resolved. Nevertheless, the squeezing energies for the first two coefficients establish the first two points anywhere near the Planck curve of the Unruh effect ever recorded (Fig.~\ref{planckcurve})

\section{Comments}

We have developed a theory that has revealed the classical root of the Unruh effect as the correlation of noise in space and time. We have demonstrated aspects of this theory in a simple laboratory experiment where we observed the squeezing of noise (Fig.~\ref{results}). The excess energy of this noise lies near  the ideal Planck curve of the Unruh effect for the first two measurable Fourier coefficients (Fig.~\ref{planckcurve}). The experiment proves that the effect is robust, even in the presence of experimental imperfections (Fig.~\ref{standingwaves}). 

Apart from the first experimental demonstration of a phenomenon in analogy to the Unruh effect, our classical analogue may also shed light on some of the more speculative facets of the effect. One may view the Unruh effect as a manifestation of the quantum vacuum as a physical substance: the quantum vacuum appears as the modern ether. One may also view it as a manifestation of inertia, distinguishing between uniform, inertial motion and accelerated, non--inertial motion. In resisting acceleration, the Unruh effect may explain deviations from acceleration that mimic hypothetical dark matter \cite{Milgrom}. Our classical analogue may show how to generalize this idea to trajectories of non--uniform accelerations. Here a straightforward extension of the quantum result is difficult, but our classical concepts still hold. 

The classical analogue of the Unruh effect may also serve in Jacobson's thermodynamic derivation \cite{Jacobson} of Einstein's equations of gravity \cite{GR}. Like Bekenstein's black--hole thermodynamics \cite{Bekenstein} that assigns an entropy to the area of the event horizon of the black hole with the Hawking temperature \cite{Hawking} as thermodynamic temperature, Jacobson assigned an entropy to any causal horizon with the Unruh temperature as thermodynamic temperature, and derived \cite{Jacobson} from these assumptions Einstein's field equations \cite{GR}. There both the entropy and the temperature carry $\hbar$'s that cancel each other. Our findings imply that the entire argument can be made classical. 

Note that Jacobson's thermodynamic derivation \cite{Jacobson} establishes an alternative to the usual derivation of Einstein's equations from the principle of least action \cite{GR}. In our opinion \cite{UL} the action principle gives the strongest argument in favor of the existence of a quantum theory of gravity, because action principles normally arise due to the quantum interference of paths or field configurations. Jacobson's derivation, combined with the classical Unruh effect, opens another, equally credible route to Einstein's classical theory of gravity \cite{GR}. On quantum gravity, it puts a question mark. 

\acknowledgments

We thank 
Rachel Bruch and
Mordehai Milgrom
for stimulating discussions. 
Our work was supported by the AXA research fund and LABEX WIFI (ANR-10-LABX-24) within the French Program ÒInvestments for the FutureÓ (ANR-10- IDEX-0001-02 PSL), the European Research Council and the Israel Science Foundation, a research grant from Mr. and Mrs. Louis Rosenmayer and from Mr. and Mrs. James Nathan, and the Murray B. Koffler Professorial Chair. Itay Griniasty is grateful to the Azrieli Foundation for the award of an Azrieli Fellowship.

\appendix

\section{Half-odd Fourier transformation}

The main difficulty of the data analysis for our --- and probably all other experimental attempts to measure the Unruh effect --- comes from the extreme time dilatation experienced by the accelerated observer. The laboratory time $t$ along the Rindler trajectory (\ref{rindler}) depends exponentially on the proper time (\ref{tau}) for large $\eta$, as $\sinh\eta\sim\mathrm{e}^\eta/2$. So in order to resolve the Planck spectrum, an exponentially large time is required (but thanks to the mirrors not an exponentially large lab space --- Fig.~\ref{mirrors}). One resolves the Planck spectrum if the characteristic factor $\mathrm{e}^{-\pi\nu}$ is resolved between the Fourier--transformed modes and the Fourier transforms of their complex conjugates. For achieving this, the resolution $\Delta\nu$ must be in the order of 
\begin{equation}
\Delta\nu = \frac{1}{2\pi} \,.
\label{delta}
\end{equation}
We obtain from the time--frequency uncertainty relation, $\Delta\nu\,\Delta\eta\sim1$, that $\Delta\eta\sim 2\pi$, which sets the minimal time window required for measuring the Planck spectrum. 

Suppose a signal along the trajectory of the accelerated observer is detected. One needs to Fourier transform and possibly filter this signal. We assume that the signal is multiplied with a filter function $F$ that describes both the finite observation time and the filtering:
\begin{equation}
A_\mathrm{F} = F(\eta)\,A(\eta) \,.
\label{filter}
\end{equation}
In the Fourier transform, $F$ appears as the convolution 
\begin{equation}
\widetilde{A_\mathrm{F}} = \frac{1}{2\pi}\int_{-\infty}^{+\infty}\widetilde{F}(\mu)\,\widetilde{A}(\nu-\mu) \,\mathrm{d}\mu\,.
\label{convolution}
\end{equation}
The most efficient way of taking data is without filtering at all:
\begin{equation}
F(\eta) = \Theta(\eta-\Delta\eta)\,\Theta(\Delta\eta-\eta)
\end{equation}
where $F$ only reflects the finite observation time we put to the minimal
\begin{equation}
\Delta\eta = \frac{1}{\Delta\nu} = 2\pi\,.
\label{deltaeta}
\end{equation}
However, avoiding filtering completely produces a problem: the Fourier transform of the finite observation window contains long, oscillatory wings:
\begin{equation}
\widetilde{F} = \frac{2\sin(\nu/\Delta\nu)}{\nu} \,.
\label{sinc}
\end{equation}
Furthermore, according to Eq.~(\ref{aktilde}), each Fourier--transformed mode has a pole at $\nu=0$. The convolution of the wings of the Fourier--transformed filter function with the pole completely obscures the Planckian relationship of Eq.~(\ref{akctilde}), unless the pole contribution vanishes.

Consider a single pole at $\nu=0$; imagine that $\widetilde{A}$ in the convolution (\ref{convolution}) is replaced by the pole. In this case the convolution integral takes the shape of the Hilbert transform \cite{Ablowitz} (Kramers-Kronig relation)
\begin{equation}
\mathrm{Re} f = \frac{1}{\pi} \dashint_{-\infty}^{+\infty} \frac{\mathrm{Im} f(\mu)}{\nu-\mu}\,\mathrm{d}\mu
\label{hilbert}
\end{equation}
for complex functions $f$ analytic on the upper half plane. Such a function is $(2/\nu)\exp(\mathrm{i}\nu/\Delta\nu)$ with the desired imaginary part (\ref{sinc}) and the real part 
\begin{equation}
\mathrm{Re} f = \frac{2\cos(\nu/\Delta\nu)}{\nu} \,.
\end{equation}
The real part, and hence the convolution of the pole, vanishes for
\begin{equation}
\nu = \frac{2n+1}{2}\,\pi\Delta\nu  = \frac{2n+1}{4} \quad\mbox{with}\quad n \in \mathbb{N} \,.
\label{nu}
\end{equation}
For filtering out the pole one should thus use finite Fourier analysis at {\it half odd integers} --- just between the usual Fourier components of periodic functions.


\begin{thebibliography}{99}

\bibitem{Unruh} 
W. G. Unruh,
Phys. Rev. D {\bf 14}, 870 (1976).

\bibitem{Fulling}
S. A. Fulling, 
Phys. Rev. D {\bf 7}, 2850 (1973).

\bibitem{Davies}
P. C. W. Davies,
J. Phys. A {\bf 8}, 609 (1975).

\bibitem{Bekenstein}
J. D. Bekenstein,
Phys. Rev. D {\bf 7}, 2333 (1973).

\bibitem{Hawking}
S. W. Hawking, 
Nature {\bf 248}, 30 (1974).

\bibitem{Crispino}
See e.g. L. C. B. Crispino, A. Higuchi, and G. E. A. Matsas, 
Rev. Mod. Phys. {\bf 80}, 787 (2008).

\bibitem{Analogues}
W. G. Unruh, Phys. Rev. Lett. {\bf 46}, 1351 (1981);
G. Volovik, 
{\it The Universe in a Helium Droplet}
(Oxford University Press, Oxford, 2003);
C. Barcelo, S. Liberati and M. Visser, Living Rev. Relativity {\bf 8}, 12 (2005);
W. G. Unruh and R. Sch\"utzhold (eds.)
{\it Quantum Analogues: From Phase Transitions to Black Holes and Cosmology}
(Springer, Berlin, 2007);
D. Faccio, F. Belgiorno, S. Cacciatori, V. Gorini, S. Liberati, and U. Moschella (eds.),
{\it Analogue Gravity Phenomenology: Analogue Spacetimes and Horizons, from Theory to Experiment},
Lecture Notes in Physics {\bf 870}
(Springer, Cham, 2013).

\bibitem{Retzker}
A. Retzker, J. I. Cirac, M. B. Plenio, and B. Reznik.
Phys. Rev. Lett. {\bf 101}, 110402 (2008).

\bibitem{Graphene}
A. Iorio and G. Lambiase, 
Phys. Lett. B {\bf 716}, 334 (2012);
Phys. Rev. D {\bf 90}, 025006 (2014);
M. Cveti\v{c} and G.W Gibbons, Ann. Phys. (N.Y.) {\bf 327}, 2617 (2012).

\bibitem{Needham}
T. Needham,
{\it Visual Complex Analysis}
(Clarendon Press, Oxford, 2002).

\bibitem{Lewenstein}
J. Rodr'guez-Laguna, L. Tarruell, M. Lewenstein, and A. Celi,
Phys. Rev. A {\bf 95}, 013627 (2017).

\bibitem{Takagi}
In the Unruh effect in odd space-time dimensions (even spatial dimensions), bosons appears as fermions and vice versa, see
S. Takagi,
Prog. Theor. Phys. Suppl. {\bf 88}, 1 (1986);
this is also true for Dirac electrons in the Beltrami trumpet \cite{Needham} made of graphene of Ref.~\cite{Graphene}.

\bibitem{Boyer}
T. H. Boyer,
Phys. Rev. D {\bf 29}, 1089 (1984).

\bibitem{Higuchi}
A. Higuchi and G. E. A. Matsas,
Phys. Rev. D {\bf 48}, 689 (1993).

\bibitem{Pauri}
M. Pauri and M. Vallisneri,
Found. Phys. {\bf 29}, 1499 (1999).

\bibitem{Water}
This is inspired by experiments for measuring the classical analogue of Hawking radiation with water waves:
S. Weinfurtner, E. W. Tedford, M. C. J. Penrice, W. G. Unruh, and G. A. Lawrence,
Phys. Rev. Lett. {\bf 106}, 021302 (2011);
L-P. Euv\'{e}, F. Michel, R. Parentani, and G. Rousseaux,
Phys. Rev. D {\bf 91}, 024020 (2015).

\bibitem{Germain}
L.-P. Euv\'{e}, F. Michel, R. Parentani, T. G. Philbin, and G. Rousseaux, 
Phys. Rev. Lett. {\bf 117}, 121301 (2016).

\bibitem{Leo}
U. Leonhardt, 
{\it Measuring the Quantum State of Light}, 
(Cambridge University Press, Cambridge, 1997).

\bibitem{Raymer}
D. T. Smithey, M. Beck, M. G. Raymer, and A. Faridani,
Phys. Rev. Lett. {\bf 70}, 1244 (1993).

\bibitem{UnruhEntanglement}
S. Massar and P. Spindel,
Phys. Rev. D {\bf 74}, 085031 (2006);
see also
B. Reznik,
Found. Phys. {\bf 33}, 167 (2003);
B. Reznik, A. Retzker, and J. Silman,
Phys. Rev. A {\bf 71}, 042104 (2005).

\bibitem{LeoBook}
U. Leonhardt, 
{\it Essential Quantum Optics: From Quantum Measurements to Black Holes}, 
(Cambridge University Press, Cambridge, 2010).

\bibitem{Rindler}
W. Rindler,
Am. J. Phys. {\bf 34}, 1174 (1966).

\bibitem{Alsing}
See also P. M. Alsing and P. W. Milonni,
Am. J. Phys. {\bf 72}, 1524 (2004).

\bibitem{Erdelyi}
A. Erd\'{e}lyi, W. Magnus, F. Oberhettinger, and F. G. Tricomi,
{\it Higher Transcendental Functions}
(McGraw-Hill, New York, 1981).

\bibitem{Remark}
One also sees this from the stationary phases in the Fourier integrals (\ref{f1}) and (\ref{f2}). The phase $\varphi = \pm k\xi \mathrm{e}^{\mp\eta}+\nu\mu$ of integral (\ref{f1}) becomes stationary $(\mathrm{d}\varphi/\mathrm{d}\eta =0$) for one point $\eta$ at the real axis, but not the phase $\varphi_* = \mp k\xi \mathrm{e}^{\mp\eta}+\nu\mu$ of integral (\ref{f2}). However, by going to $\eta+\mathrm{i}\pi$ the phase $\varphi_*$ becomes the same as $\varphi$, apart from $\mathrm{i}\pi\nu$ that results in the exponential factor (\ref{exp}).

\bibitem{Phase}
One obtains from the method of stationary phase \cite{Remark} $\phi\sim\nu\ln\nu-\nu-\pi/4$ for large $\nu$. 

\bibitem{Moments}
Expanding the sine in Eq.~(\ref{amplitude}) into exponentials one gets for $\langle q(\nu_1) q(\nu_2) \rangle$ and also for $\langle p(\nu_1) p(\nu_2) \rangle$ the expression
$$
\begin{aligned}
\frac{I}{4\pi}&\int_0^\infty \left(\mathrm{e}^{\mathrm{i}(\nu_1-\nu_2)\ln k\xi}+\mathrm{e}^{\mathrm{i}(\nu_2-\nu_1)\ln k\xi} -\mathrm{e}^{\mathrm{i}(\nu_1+\nu_2)\ln k\xi}\right.\\
&\quad\quad\quad\left.-\mathrm{e}^{-\mathrm{i}(\nu_1+\nu_1)\ln k\xi}\right)\frac{\mathrm{d}k}{k}\\
&= \frac{I}{2}\,\delta(\nu_1-\nu_2) - \frac{I}{2}\,\delta(\nu_1+\nu_2) 
\end{aligned}
$$
that gives Eq.~(\ref{numoments}) since $\nu>0$.

\bibitem{Faraday}
M. Faraday,
Phil. Trans. Roy. Soc. London {\bf 121}, 299 (1831);
S. Douady, J. Fluid Mech. {\bf 221}, 383 (1990).

\bibitem{Height}
H. Murase, Proc. 3rd Int. Conf. Computer Vision, pp. 313-317 (1990); 
F. Moisy, M. Rabaud, and K. Salsac, 
Exp. Fluids {\bf 46}, 1021 (2009).

\bibitem{Milgrom}
M. Milgrom,
Phys. Lett. A {\bf 253}, 273 (1999).

\bibitem{Jacobson}
T. Jacobson,
Phys. Rev. Lett. {\bf 75}, 1260 (1995).

\bibitem{GR}
L. D. Landau and E. M. Lifshitz,
{\it The Classical Theory of Fields}
(Butterworth-Heinemann, Amsterdam, 2003);
P. A. M. Dirac,
{\it General Theory of Relativity}
(Princeton University Press, Princeton, 1996).

\bibitem{UL}
U. Leonhardt,
in M. McCall {\it et al.},
J. Opt. B (in press). 

\bibitem{Ablowitz}
M. J. Ablowitz and A. S. Fokas,
{\it Complex Variables: Introduction and Applications}
(Cambridge University Press, Cambridge, 2003).

\end{thebibliography}
\end{document}